%
\let\useblackboard=\iftrue
%
\let\useblackboard=\iffalse
%
%
\input harvmac.tex
%
\input epsf.tex
\ifx\epsfbox\UnDeFiNeD\message{(NO epsf.tex, FIGURES WILL BE
IGNORED)}
\def\figin#1{\vskip2in}
\else\message{(FIGURES WILL BE INCLUDED)}\def\figin#1{#1}\fi
\def\ifig#1#2#3{\xdef#1{fig.~\the\figno}
\midinsert{\centerline{\figin{#3}}%
\smallskip\centerline{\vbox{\baselineskip12pt
\advance\hsize by -1truein\noindent{\bf Fig.~\the\figno:} #2}}
\bigskip}\endinsert\global\advance\figno by1}
\noblackbox
\baselineskip=12pt
\useblackboard
\message{If you do not have msbm (blackboard bold) fonts,}
\message{change the option at the top of the tex file.}

\font\blackboard=msbm10 scaled \magstep1
\font\blackboards=msbm7
\font\blackboardss=msbm5
\textfont\black=\blackboard
\scriptfont\black=\blackboards
\scriptscriptfont\black=\blackboardss

\else

\fi

\def\lsim{\mathrel{\lower4pt\hbox{$\sim$}}
\hskip-12.5pt\raise1.6pt\hbox{$<$}\;}

\def\gsim{\mathrel{\lower4pt\hbox{$\sim$}}
\hskip-12.5pt\raise1.6pt\hbox{$>$}\;}

\def\yboxit#1#2{\vbox{\hrule height #1 \hbox{\vrule width #1
\vbox{#2}\vrule width #1 }\hrule height #1 }}
\def\fillbox#1{\hbox to #1{\vbox to #1{\vfil}\hfil}}
\def\ybox{{\lower 1.3pt \yboxit{0.4pt}{\fillbox{8pt}}\hskip-0.2pt}}

\def\comments#1{}

\def\CN{{\cal N}}

\def\II{\relax{I\kern-.07em I}}

\def\IZ{\relax\ifmmode\mathchoice
{\hbox{\cmss Z\kern-.4em Z}}{\hbox{\cmss Z\kern-.4em Z}}
{\lower.9pt\hbox{\cmsss Z\kern-.4em Z}}
{\lower1.2pt\hbox{\cmsss Z\kern-.4em Z}}\else{\cmss Z\kern-.4em
Z}\fi}

\def\nfour{$\CN \!=\! 4$}

\def\ntwo{$\CN \!=\! 2$}

%
%

%
%
\lref\seibergi{S. Minwalla, M. Van Raamsdonk and N. Seiberg,
``Noncommutative Perturbative Dynamics'', hep-th/9912072.}
\lref\haya{ M. Hayakawa, ``Perturbative analysis on infrared and ultraviolet aspects of noncommutative QED on $R^4$'', hep-th/9912167.}
\lref\grosse{H. Grosse, T. Krajewski and R. Wulkenhaar, ``Renormalization of noncommutative Yang-Mills theories: A simple example'', hep-th/0001182.}
\lref\are{I. Ya. Aref'eva, D. M. Belov, A. S.Koshelev, ``A Note on UV/IR for Noncommutative Complex Scalar Field'', hep-th/0001215.}
\lref\sussk{A. Matusis, L. Susskind and N. Toumbas, ``The IR/UV Connection in the Non-Commutative Gauge Theories'', hep-th/0002075}
\lref\chon{Chong-Sun Chu, ``Induced Chern-Simons and WZW action in Noncommutative Spacetime'', hep-th/0003007.}
\lref\seibergone{M.~Van Raamsdonk and N.~Seiberg,
``Comments on noncommutative perturbative dynamics,''
JHEP {\bf 0003}, 035 (2000), hep-th/0002186.}
\lref\andr{O. Andreev and H. Dorn, ``Diagrams of Noncommutative Phi-Three Theory from String Theory'', hep-th/0003113.} 
\lref\camp{B. A. Campbell, K. Kaminsky, ``Noncommutative Field Theory and Spontaneous Symmetry Breaking'', hep-th/0003137.}
\lref\kiem{Y. Kiem and S. Lee ``UV/IR Mixing in Noncommutative Field Theory via Open String Loops'',  hep-th/0003145.}
\lref\aredos{I. Ya. Aref'eva, D. M. Belov, A. S. Koshelev and O. A. Rytchkov, ``UV/IR Mixing for Noncommutative Complex Scalar Field Theory, II (Interaction with Gauge Fields)'', hep-th/0003176.}
\lref\bilal{A. Bilal, Chong-Sun Chu and R. Russo, ``String Theory and Noncommutative Field Theories at One Loop'', hep-th/0003180.}
\lref\jgomis{J. Gomis, M. Kleban, T. Mehen, M. Rangamani and S. Shenker ``Noncommutative Gauge Dynamics From The String Worldsheet'', hep-th/0003215.}
\lref\raj{A. Rajaraman and M. Rozali, ``Noncommutative Gauge Theory, Divergences and Closed Strings'',  hep-th/0003227.}
\lref\liu{H. Liu and J. Michelson, ``Stretched Strings in Noncommutative Field Theory'', hep-th/0004013.}
\lref\bich{A. A. Bichl, J. M. Grimstrup, V. Putz and M. Schweda, ``Perturbative Chern-Simons Theory on Noncommutative $R^3$'', hep-th/0004071.}
\lref\frede{F. Zamora  `` On the Operator Product Expansion in Noncommutative Quantum Field Theory'', hep-th/0004085.}
\lref\gll{J. Gomis, K. Landsteiner and E. Lopez, ``Non-Relativistic Non-Commutative Field Theory and UV/IR Mixing'', hep-th/0004115.}
\lref\russo{Chong-Sun Chu, R. Russo and S. Sciuto, ``Multiloop String Amplitudes with B-Field and Noncommutative QFT'', hep-th/0004183.}
\lref\meh{J. Gomis and T. Mehen, ``Space-Time Noncommutative Field Theories and Unitarity'', hep-th/0005129.}
\lref\armoni{A. Armoni, ``Comments on Perturbative Dynamics of Non-Commutative Yang-Mills Theory'', hep-th/0005208.}
\lref\bak{D. Bak, S. K. Kim, Kwang-Sup Soh and J. H. Yee, ``Exact Wavefunctions in a Noncommutative Field Theory'', hep-th/0005253.}
\lref\bras{H. O. Girotti, M. Gomes, V. O. Rivelles and A. J. da Silva, ``A Consistent Noncommutative Field Theory: the Wess-Zumino Model'', hep-th/0005272.}
\lref\chaudhuri{S.~Chaudhuri and E.~G.~Novak,
``Effective string tension and renormalizability: String theory in a  noncommutative space,''
JHEP {\bf 0008}, 027 (2000), hep-th/0006014.
}
\lref\mg{G. Arcioni, J. L. F. Barbon, Joaquim Gomis and M. A. Vazquez-Mozo,
``On the stringy nature of winding modes in noncommutative thermal field theories'',
hep-th/0004080.}
\lref\fischleri{W. Fischler, Joaquim Gomis, E. Gorbatov, A. Kashani-Poor, S. Paban and P. Pouliot, `` Evidence for Winding States in Noncommutative Quantum Field Theory'', JHEP 0005 (2000) 024,  hep-th/0002067.}
\lref\fischlerii{ W. Fischler, E. Gorbatov, A. Kashani-Poor, R. McNees, S. Paban and P. Pouliot, ``The Interplay Between $\theta$ and T'', hep-th/0003216.}
\lref\kapusta{J. Kapusta, ``Finite-temperature field theory'', 
Cambridge Univ. Press 1989.}
\lref\whitham{G. B. Whitham, ``Linear and Nonlinear Waves'', John Wiley \& Sons 1974.}
\lref\sexl{R. Sexl and H. Urbantke, ``Relativit\"at, Gruppen, Teilchen'',
 Springer 1976.}
\lref\brillouin{L. Brillouin, ``Wave propagation and Group Velocity'',
New York Academic Press 1960.}
\lref\gubser{S.~S.~Gubser and S.~L.~Sondhi,
``Phase structure of non-commutative scalar field theories,''
hep-th/0006119.}
\lref\miguel{G. Arcioni, M. A. Vazquez-Mozo, ``Thermal effects in perturbative noncommutative gauge theories'', JHEP 0001 (2000) 028, hep-th/9912140.}
\lref\weldon{H. E. Haber, H. A. Weldon, ``Finite Temperature Symmetry Breaking as Bose-Einstein Condensation'', Phys.Rev.D25:502,1982.}
\lref\finite{M. T. Grisaru and W. Siegel, ``Improved Methods for Supergraphs'',Nucl.Phys.B159 (1979) 429.}

%
\Title{ \vbox{\baselineskip12pt\hbox{hep-th/0006210}
\hbox{CERN-TH-2000-173}
}}
{\vbox{
\centerline{Excitations in Hot Non-Commutative Theories} }}

\centerline{Karl Landsteiner, Esperanza Lopez and Michel H.G. Tytgat}
\medskip
\centerline{Theory Division CERN}
\centerline{CH-1211 Geneva 23}
\centerline{Switzerland}
\medskip
\centerline{\tt Karl.Landsteiner@cern.ch}
\centerline{\tt Esperanza.Lopez@cern.ch}
\centerline{\tt Michel.Tytgat@cern.ch}

\vskip15mm

\centerline{\bf Abstract}
We study the dispersion relation for scalar excitations  in supersymmetric, 
non-commutative theories at finite temperature.
 In   
${\cal N}\!=\!4$ Yang-Mills the low momenta modes have superluminous group 
velocity.  In the massless Wess-Zumino model the minimum of the 
dispersion relation is at
non zero momentum  for temperatures above  
 $T_0 \approx (g \theta)^{-{1 \over 2}}$.
We briefly comment on ${\cal N}\!=\!2$ Yang-Mills at finite
density.
\baselineskip=14pt

\vfill
\Date{\vbox{\hbox{\sl June 2000}}}

\newsec{Introduction}
Field theories on non-commutative space have interesting properties.
In particular, infrared divergences appear whose origin are ultraviolet 
degrees of freedom circulating in loops. This phenomenon implies a
surprising mixing between the ultraviolet and infrared degrees of
freedom of the theory \seibergi. The study of infrared singularities
appearing in perturbation theory 
has been an active field of research recently 
\refs{\haya \grosse \are \sussk \chon \seibergone \andr \camp \kiem \aredos \bilal \jgomis \raj \liu \bich \frede \gll \russo \meh \armoni \bak  \bras {--} \chaudhuri }.

In this paper we will study non-commutative field theories
at finite temperature and density. In particular we will study the
dispersion relation of scalar fields.
The behaviour of hot non-commutative field theories 
has been investigated in \refs{\miguel \fischleri \fischlerii {--} \mg}. At high temperature, when the 
characteristic thermal 
wavelength of the particles becomes smaller than
the radius of the Moyal cell, $T \gsim \theta^{-1/2}$, one observes a 
suppression of the contribution from non-planar graphs with respect to the planar 
ones. This suggests that there is a 
reduction of the number of degrees of freedom running in non-planar graphs 
\fischleri \fischlerii.

The study of the dispersion relation is relevant because it directly addresses 
the question of how the physical degrees of freedom of non-commutative field 
theories differ from their commutative counterparts.
It is by now well-known that at zero temperature and density
the one loop dispersion relation gets modified by terms showing a
pole like infrared singularity \seibergi which is 
typically of the form
\eqn\dispersionvacuum{
\omega^2 = p^2 + m^2 + {c\over \vert\tilde{p}^2\vert} \,.} 
Here we used the notation $\tilde{p}^\mu = \theta^{\mu\nu} p_\nu$
and $c$ is a model dependent constant, $c = {\cal O}(g^2)$.  
These infrared poles have their origin in 
the regularization of the quadratic divergences in non-planar
diagrams by the Moyal phases. If the external momentum flowing into
the diagram vanishes, the original divergence is recovered but disguised
as an infrared divergence.
In supersymmetric theories  the quadratic divergences are cancelled 
by Bose-Fermi degeneracy ($c = 0$).
In a heat bath supersymmetry is broken.
The one loop  dispersion relations
are then bound to be non-trivial while the infrared behaviour is 
under control in perturbation theory.
It is the aim of this paper to investigate
these modifications and their consequences
in non-commutative field theories.

In section two, we discuss the scalar one loop dispersion relation  in 
\nfour\ non-commutative Yang-Mills at finite temperature.
We find that for all temperatures
the group velocity exceeds the speed of light for soft 
momenta in the non-commutative direction.
We study the problem of wave propagation
in the limit of small momenta.
These modes obey the dispersion relation of the linearised Korteweg-deVries
equation. The wavefront of a disturbance travels faster than light. It is perhaps surprising that
the speed of light is no more a barrier. This is per se not inconsistent
since Lorentz symmetry is absent in non-commutative space.

In section three, we 
analyse the non-commutative version of the Wess Zumino model.
For sufficiently high temperature a dip appears in the dispersion
relation. 
More precisely, the lowest energy mode has non-vanishing
momentum in the non-commutative directions. 
At
high temperatures and small momenta, the group velocity again
exceeds the speed of light. 
We furthermore argue that the dip makes Bose-Einstein condensation
impossible. 

In section four we analyse the effects of non-commutativity on field
theories at finite density. We will investigate
\ntwo\ supersymmetric gauge theory with a chemical potential $\mu$ for the
gauginos. 
The dispersion relation for the scalars at one loop
is similar to the one encountered in section two. The role of temperature
is now played by the chemical potential.

\newsec{\nfour\ Yang-Mills}

We will work in four-dimensions and consider only 
space non-commutativity. Without loss of generality we 
take
\eqn\com{
[ x^{\mu},x^{\nu} ]= i \theta^{\mu \nu}
} 
with $\theta^{12} \! =\!-\theta^{21}\!=\!\theta$ and 
$\theta^{\mu \nu}\!=\!0$ otherwise. The algebra of functions on 
this space is defined through the star product
\eqn\star{
(f*g)(x) := \lim_{y\rightarrow x}
e^{{i \over 2}\theta^{\mu\nu} \partial^x_\mu\partial^y_\nu}f(x) g(y) \,
}
where here $x^\mu$ are taken to be ordinary c-numbers. 

The non-commutative version of \nfour\ Yang-Mills \fischlerii\ is obtained by 
substituting the star product in commutators, or equivalently,
replacing commutators by the so-called Moyal bracket
\eqn\moyal{ \left\{ f , g \right\} = (f*g)(x) - (g*f)(x)\,.}
In a non-commutative gauge theory with gauge group $U(N)$ the  
Moyal phases derived from the star product cancel
in the $SU(N)$ part at the one loop level \seibergi\ \armoni.
In others words there are no non-planar contributions from particles 
in the adjoint of $SU(N)$ running in the loop and the only non-planar one loop
graph stems from the $U(1)$ factor. For this reason we limit ourselves to 
the study of a $U(1)$ \nfour\ gauge theory. 
The spectrum of the theory consists of six scalars, four Majorana Fermions
and a vector field. The corresponding Feynman rules can be found in 
\haya\ \sussk.

We will study the dispersion relation
of the \nfour\ scalars at finite temperature and one loop level.
We will comment later on the modifications for fermions and gauge bosons. 
The scalar self-energy is given by
\eqn\propnfour{\Sigma_T = 32 g^2 \int
 {d^3k \over (2\pi)^3} {\sin^2 {\tilde{p}\cdot k\over  2} \over k} 
\left( n_B(k) + n_F(k) \right) + 4 g^2 P^2 \bar{\Sigma} \,.} 
We use uppercases 
for the four momenta and lowercases for the spatial components, $P^2 = 
p_0^2 - p^2$. Also $\tilde p k \equiv \theta(p_1 k_2 - p_2 k_1)$
and we refer to  momenta along the non-commutative directions as transverse
momenta, with $\tilde p^2 = \theta^2(p_1^2 + p_2^2) = \theta^2 p_\perp^2$. 

The first term in \propnfour\ vanishes at $T=0$ because of 
supersymmetry.
The second term contributes 
to the finite temperature wave-function renormalization of the scalar field. 
It affects the position of the pole only to ${\cal O}(g^4)$ and we
will drop it in the sequel. A subtlety should be mentioned. 
The zero temperature planar contribution to $\bar \Sigma$ is 
typically logarithmically divergent. As a consequence the non-planar contribution 
can give raise to an infrared logarithmic singularity. For a
massless theory we will have
\eqn\loga{
{\bar \Sigma}_{T=0}^{np} \sim g^2 P^2 {\rm log} 
{1 \over {\tilde p}^2 P^2} \, .
} 
If such a term is present \foot{This term is gauge dependent. 
In ordinary \nfour\ gauge theories both quadratic and logarithmic
divergences are absent if a convenient gauge is
chosen \finite. One can speculate that in the non-commutative version
of \nfour\ there must also exist a gauge in which the theory
is smooth in the infrared \sussk.}, the perturbative expansion  
breaks down for non-perturbatively small transverse momenta, 
${\cal O}(e^{-{1 \over g^2}})$ \seibergi. 
The phenomena we will find in this and the next sections do not
depend on entering into this region. Thus in the
following we will ignore contributions to the 
self-energy proportional to $g^2 P^2$. 

\ifig\bumps{
Dispersion relation for scalars in \nfour\ Yang-Mills for different temperatures.
The momentum $p$ is taken to lie entirely in the non-commutative directions.
The dashed line shows the light cone $\omega=p$. The dotted line
shows the momentum $p_c$ below which the group velocity 
${\partial \omega\over \partial p}$ is bigger
than one.}
{
\epsfxsize=3.8truein\epsfysize=2.3truein
\epsfbox{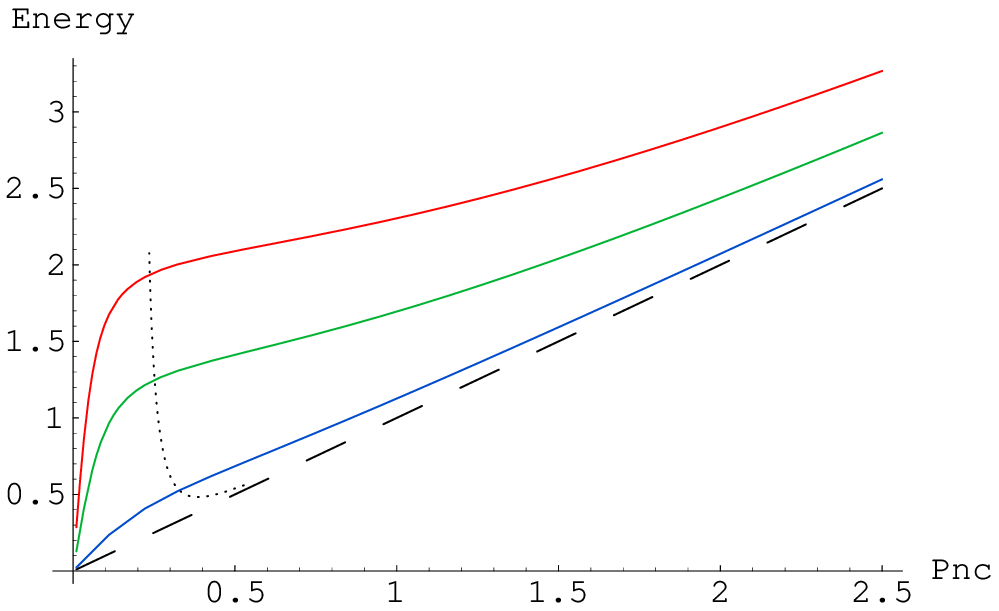}
}

Using the relation $\sin^2{\tilde{p} k\over 2} = {1\over 2} (1 - \cos\tilde p k)$ we can separate the planar and non-planar contributions to the self-energy.
The dispersion relation simply becomes
\eqn\dispersionnfour{ \omega^2 = p^2 + 2g^2 T^2 - 
{4 g^2 T\over \pi \tilde{p}} \tanh{\pi \tilde{p} T\over 2} \, ,}
where $\tilde p \equiv \vert \tilde p\vert$. 
The qualitative features of the dispersion relation,
which is  plotted in \bumps , are easy to understand. 
The hyperbolic tangent arises solely from the non-planar contribution to the 
dispersion relation.
For large transverse external momenta the non-planar contribution
is subleading with respect to the planar one,
\eqn\highpt{
\omega^2 \approx p^2 + 2 g^2 T^2 - 4 g^2 {T\over \pi \tilde p}, \;\;\;\  T\tilde p \gg 1.
}
The second term comes from the planar diagrams and gives a mass to the 
scalar excitations.
The subdominant term linear in $T$ is quite interesting. It arises solely from soft bosons in non-planar diagrams. 
These are modes with characteristic momentum $k \ll T$ and large occupation number,
$n_B \approx T/k ~\gg~1$,
\eqn\soft{
\Sigma_{np} \sim \int d^3k {1\over k} \cos{\tilde p  k} {T\over k} \sim {T\over \tilde p} \, .
}
The ultraviolet catastrophe of usual space-time does not arise as long as 
 the non-commutativity scale, or more precisely $\tilde p$, is kept non-zero. This is yet another 
manifestation of the UV/IR mixing of non-commutative field theories: to
leading order at high temperature, the non-planar contribution is effectively purely classical \fischlerii. 
At low traverse external momenta, the non-planar contribution tend to 
cancel 
the planar one,
\eqn\lowpt{
\omega^2 \approx p^2 + 2 g^2 T^2 - 2 g^2 T^2 + {g^2 \pi^2 T^4\over 6} \tilde p^2, \;\;\; T \tilde p \ll 1 \, .
}
For zero external transverse momentum the interaction switches off.
The theory
becomes a free, gapless $U(1)$ gauge theory with $\omega^2 \approx p_3^2$.

\bigskip

Let us consider now the case where the momentum lies
along the non-commutative directions, $p^2=p_\perp^2$ .
Since $\omega(0)=0$ and for large $p$, $\omega(p) \approx  \sqrt{p^2+2 g^2 T^2}$, which lies above the lightcone, 
there is a region in between with ${\partial\omega(p)\over \partial p} > 1$.
 Thus the group velocity must exceed the speed of light
for small transverse momenta\foot{By small transverse momenta we mean
small but still outside the exponentially small region where logarithmic 
terms as \loga\ could become important.}!  From \lowpt\ we find 
\eqn\ftl{
\omega^2 \approx \left(1 + {g^2 \pi^2 T^4 \theta^2\over 6} \right)\; p^2 \, .
} 
The low momentum excitations are massless, but propagate with an index of refraction $n = p/\omega$ 
 smaller than one. Because the interactions switch off at low momenta, 
we expect these modes to be long-lived.\foot{
At one loop the self-energy correction we are considering is real since
it is essentially a tadpole contribution. It depends of the external
momentum only because of the star product.}  
The dispersion relation for several temperatures is plotted in \bumps . 
The dashed line shows the light cone $\omega=p$. Below some
momentum $p_c$ the group velocity exceeds the speed of light. For high
temperature,
\eqn\pcrit{
p_c \approx \sqrt{{\sqrt{2}\over \pi \theta}g}\,.
}

Let us  emphasise that these qualitative features should be quite general and not an artifact of our one loop approximation,
as they simply arise from the fact that the theory is non-interacting at zero transverse momentum and develops a 
mass gap otherwise. Fermions and gauge bosons dispersion relations are involved at finite temperature. However, it should be clear that these particles share the same qualitative properties as the scalar modes. For instance, the Debye
mass of longitudinal gauge bosons is vanishing in the limit of zero transverse momentum.   
\foot{
The perturbative expansion is much 
better behaved than in usual thermal field theories. This is because the theory cures its infrared divergences 
by switching off the interactions at soft momenta. For instance, there is no need to resum daisy 
or Linde diagrams \kapusta\ and, to all order in perturbation theory, the free energy  is given by an expansion in
 $g^2$ \miguel. Similarly, the correction to the dispersion relation are ${\cal O}(g^4)$.} 
  
\bigskip

We now investigate  the consequences of the  dispersion relation \dispersionnfour\ for wave propagation.
Imagine that some disturbance of the scalar field is created in the thermal bath at time $t=0$.   
An elementary solution can be written in the form
\eqn\wave{ \Phi(x,t) = \int_{-\infty}^{\infty} \left( F(k) 
e^{i(k x-\omega t)} + F^*(k) e^{-i(k x-\omega t)}\right)dk \, ,}
where $F(k)$ is determined by an initial condition at $t=0$. 
To simplify matters we will consider only a one dimensional 
problem with momentum pointing in a noncommutative direction.\foot{
For large times $t$ we can apply a stationary wave argument and find
that $\Phi$ at the point $x$ receives a contribution from the
modes $k=\kappa$ satisfying $\omega^\prime(\kappa) = {x\over t}$.
The asymptotic behaviour of the wave is given by \whitham\
\eqn\wavei{\Phi = 2 Re \left( F(k) \sqrt{{2\pi\over t
|\omega^{\prime\prime}(\kappa)|}} \exp\left(i k x - i \omega(\kappa) t - i{\pi\over4}sgn(\omega^{\prime\prime}(\kappa))
\right)\right)\,.}
If $\omega^{\prime\prime}(\kappa) =0$ we have to go further in the Taylor
expansion around the point of stationary phase.}
The fastest moving modes are the ones with longest wavelength. These are also the modes 
which are long lived in the thermal bath. For these
 it is possible to obtain the exact asymptotic behaviour by noting that
the dispersion relation around $k=0$ is 
\eqn\lkdv{ \omega(k) = c_0 k - \gamma k^3 + O(k^5)\,,}
with $c_0 = \sqrt{1 + {g^2 \pi^2 T^4 \theta^2 \over 6}}$ and $\gamma = {g^2 \pi^4 \theta^4 T^6\over 120 c_0}$.
This is the dispersion relation of the linearised Korteweg-deVries equation whose solution
 is expressed in terms of the Airy function $Ai(z)$. If we further
use $F(\kappa) \approx F(0)$ for small $k$ we can express the solution
for the head of the wavetrain by
\eqn\airy{ \Phi = {F(0)\over 2(3\gamma t)^{1\over3}} Ai\left(
{x-c_0 t\over (3\gamma t)^{1\over3}}\right) \,.}
The Airy function has oscillatory behaviour for negative argument
and decays exponentially for positive argument. Thus the wavetrain decays
exponentially ahead of $x=c_0 t$. Behind the wave becomes oscillatory
and the Airy function matches onto the expression \wavei. 
In between the oscillatory region and the exponential decay
there is a transition region of width proportional to $(\gamma t)^{1\over 3}$
around $x=c_0 t$. In this region the wavetrain has its first crest which
therefore is moving with a velocity approximately given by $c_0$.

\ifig\firstcrest{
The movement of the first crest of a wavetrain. 
We have chosen $g=0.1$, $T=4$ and $\theta=1$. With this
choice $c_o\approx 2.2$. The crest moves with approximate speed
$c_o$.}
{
\epsfxsize=3.8truein\epsfysize=2.4truein
\epsfbox{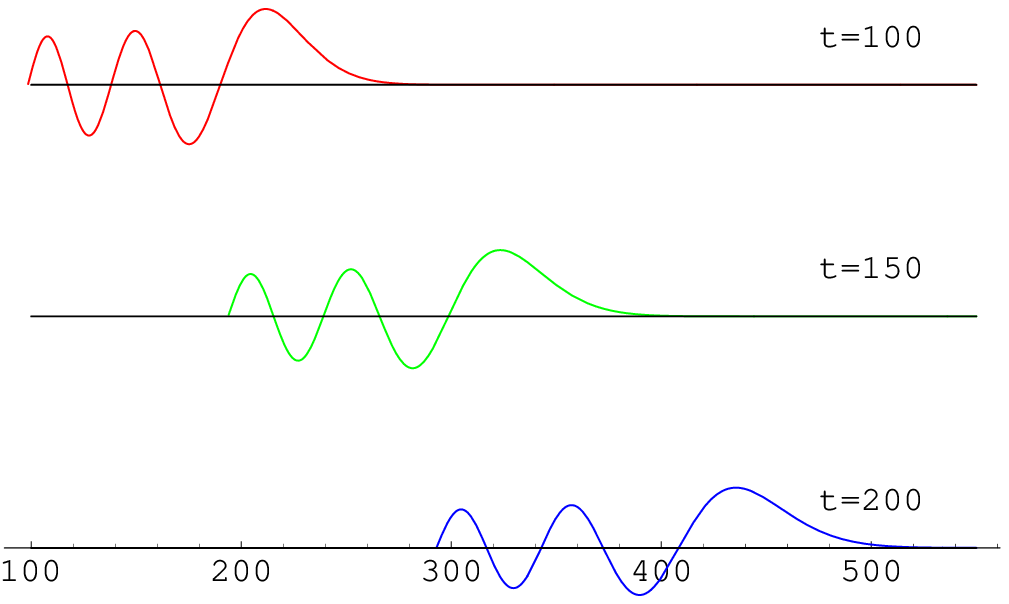}
}
A condition as to when the first crest of the Airy function is
a good approximation of the asymptotic behaviour of the wavetrain
can be obtained as follows. The spread of the first crest gives us
a length scale which should correspond to a wavelength lying in
the range where \lkdv\ is valid.
This is the case if the term of order $O(k^5)$ is small
against the first two. This typically happens for 
$k<{1\over \theta T}$. That the first crest can be built up by 
the corresponding modes implies then $t>{(\theta T)^3\over \gamma}$\foot{For very long times instabilities of the scalar excitations
arising at two loop will effect the behaviour of the wavefront.
Although we do not know these effects in more detail we expect
that there is a regime where our one loop expression is a
good approximation for long enough times.}.  
The movement of the first crest is shown in \firstcrest. 
Note that the behaviour at large $x$ is independent 
of the precise
form of the initial conditions.

Group velocities faster than the speed of light do also appear in
conventional physics, e.g. it is well-known that this happens
for light propagation in media close to an absorption line. 
Since the dispersive effects are however large, 
the group velocity loses its meaning as the velocity of signal
transportation. In our case, it is interesting to notice that
as the temperature increases, not only $c_0$ but also $\gamma$ grows.
This implies that at high temperatures the soft transverse momenta
become very dispersive. In such situations it is useful to introduce the 
concept of a front
velocity which is the velocity of the head of the wavetrain \sexl.
For the propagation of light in a medium it can be shown
that this front velocity never exceeds the speed of light even if
the group velocity can be faster than the speed of light \brillouin. 
In our case the front velocity can be defined as the velocity of
the first crest of the wavetrain. According to \dispersionnfour\ and
\lkdv\ this is always bigger that the speed of light.
The advance of the first crest with respect to an imagined
light front is $(c_0-1)t$. Since its spread grows as $(\gamma t)^{1
\over 3}$, the first crest is well defined outside the lightcone
for large enough time, $t> t_0$ where $t_0=\sqrt{\gamma \over (c_0-1)^3}$.

The wavefront defined in the previous way is independent of the detailed
form of the signal. It deserves more detailed study to answer the
question of how fast can we transmit a signal. However, at least
the fraction of the energy corresponding to the first crest of the wavetrain
moves faster than the speed of light.  
Could this be a hint of violation of causality?
We think this is not the case in these theories.
Since the Lorentz symmetry $SO(1,3)$ is broken to $SO(1,1)\times SO(2)$ there
is no group theoretical reason for maximum velocity in the transverse directions. There exists a unique 
class of reference frames where the non-commutativity takes the form
$[x^1,x^2] = i \theta$. If we naively perform a Lorentz boost in the 
say $x^2$ 
direction we find that in the new coordinate system we not only have
space-space but also space-time non-commutativity. An observer being
at rest in such a reference frame could measure $[x^0,x^2] = i \theta_1$ and
$[x^1,x^2] = i \theta_2$. Since the non-commutativity involves three coordinates
now the observer could dualize the tensor to a vector and compute its
norm using the Minkowski metric.
He would find $\theta_1^2-\theta_2^2 < 0$. From
this he could conclude that there is a frame in which time commutes
with the other coordinates. Using this time coordinate he would also conclude
that energy propagation is always forward in time. Thus the notion of an
absolute, commutative time coordinate should prevent acausality to happen.
Of course this still depends on the assumption that the laws of physics
as formulated in the boosted reference frame are such that energy
propagation is always forward in commuting time. This can not be proven
at the moment since we do not know how to formulate a dynamical principle
with non-commuting time coordinates. Thus, although we can not prove
that the theory is causal, we can at least show that acausality is not
automatically implied by superluminous effects of \dispersionnfour.

\newsec{Wess-Zumino Model}

In the past section we have obtained very striking effects
associated with non-commutativity in supersymmetric gauge theories.
In these theories, the star-product enters in the interaction 
through the substitution of commutators by Moyal brackets \moyal.
We will consider now the non-commutative version of the massless 
Wess-Zumino model, which is defined by the Lagrangian \fischleri
\eqn\WZ{
{\cal L}= \partial_{\mu} \phi^{\dagger} \partial^{\mu} \phi
+ i {\bar \psi} {\bar \sigma}^{\mu} \partial_{\mu} \psi - g \psi \! 
\ast \! \psi \,
\phi -g {\bar \psi} \! \ast \! {\bar \psi} \, \phi^{\dagger} -
g^2 \phi^{\dagger}\! \ast \!\phi^{\dagger} \, \phi \! \ast \! \phi \, .
}
The interaction vertices of $U(1)$ non-commutative theory are
proportional to ${\rm sin} {{\tilde p} k \over 2}$, where $p,k$ are
momenta entering into the vertex, implying that the interaction switches 
of at low momentum. Contrary, the vertices for the Wess-Zumino model are 
proportional to ${\rm cos} {{\tilde p} k \over 2}$. The aim of this
section is to study how this difference influences the dispersion
relations.

\ifig\dips{Plot of $\omega$ as a function of $p_\perp$
for different temperatures and $p_3\!=\!0$ 
($g^2\!=\!0.1$, $\theta\!=\!1$).
For $T\!>\!T_0$ the dispersion relation develops a dip. The dotted line
locates the position of the minimum.}
{
\epsfxsize=3.8truein\epsfysize=2.4truein
\epsfbox{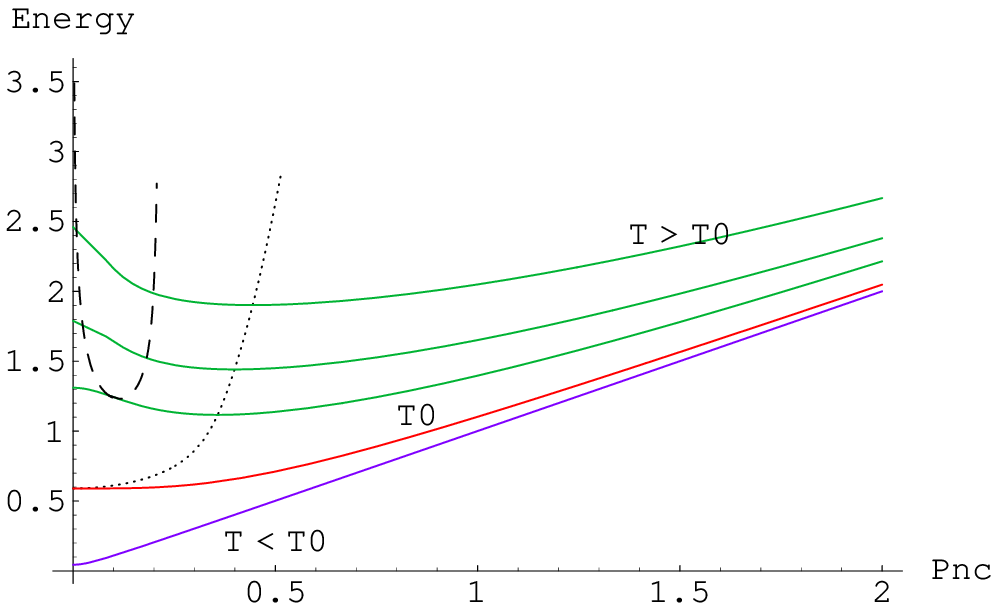}
} 

At finite temperature the one loop contribution to the scalars 
self-energy is given by
\eqn\propT{
\Sigma_T= 4 \, g^2 \! \int \! {d^3 k \over (2 \pi)^3} 
{{\rm cos}^2 {{\tilde p} k \over 2} \over k} 
\big( n_B(k) + n_F(k) \big) \,  .
}  
We have ignored terms proportional to $g^2 P^2$
for the same reasons already explained in the previous section.
We obtain the following dispersion relation
\eqn\dispersion{
\omega^2 = p^2 \, + \, {g^2 T^2 \over 4} \, + \, {g^2 T \over 2 \pi {\tilde p}} 
\, {\rm tanh} {\pi T {\tilde p} \over 2}  \, .
} 
Contrary to \dispersionnfour, the planar and non-planar corrections,
second and third term in \dispersion\ respectively, add up instead of 
subtracting. As a result all the modes down to zero momentum
acquire a mass due to thermal effects.
At small temperature the minimum is at zero momentum.
Since the non-planar contribution tends to the planar one in the small 
$\tilde p$ limit, the thermal mass is $m^2 \approx 
2 \, m^2_{planar}={g^2 T^2 \over 2}$.
At high temperature the system behaves quite differently.    
There is a temperature
\eqn\Tcrit{
T_0  = \left( {48 \over \pi^2 g^2 \theta^2} \right)^{1 \over 4} \, ,
}
such that for $T\! > \! T_0$ 
the lowest energy  mode lies at $p^{min}_\perp > 0$, as can be seen in 
\dips. For large temperatures we find
\eqn\pmin{
p_{\perp}^{min} \approx \left( {g^2 T \over 4 \pi \theta} 
\right)^{1 \over 3} \, .
} 
Temperatures bigger than $T_0$
implies $T^2 \theta \gsim  {1 \over g}$. At weak coupling, this effect is 
seen at temperatures much higher than the expected scale for the onset of 
non-commutative effects $T^2 \theta \! > \! 1$.

As $T$ increases and for $p_\perp > p_\perp^{min}$
the  thermal  mass tends to $m_{planar}^2 = {g^2 T^2\over 4}$. 
Clearly, 
because $p_\perp^{min} \propto T^{1/3}$, at sufficiently  high 
temperature the group velocity ${\partial \omega \over \partial 
p_\perp } < -1$. This is indicated in \dips\ by the region inside the 
dashed line.
As discussed in the previous section in footnote 5 we can
apply a stationary phase approximation. Note that this is valid 
whenever the imaginary
part of the dispersion relation vanishes\foot{${\partial 
\omega \over \partial p } < -1$
also happens for propagation of light in a dispersive medium close 
to an absorption line.
However the imaginary part of the dispersion relation does not vanish 
and this is the reason
why one has to apply a more accurate saddle point method \brillouin. 
In this case the saddle point
evaluation shows that the speed of light is never exceeded.}. 
This is the case to the approximation we are working.
Thus the modes with  
${\partial \omega \over \partial p_\perp } < -1$ contribute to a wave at $x = 
t \omega^\prime$ outside the lightcone. 

\bigskip 

The dip in the dispersion relation has also rather interesting consequences
for the phase structure of the theory. Consider 
for instance a gas which is
not only hot but also carries some net global charge. 
The massless Wess-Zumino model has a $U(1)$ R-symmetry under which 
the the scalar and Weyl fields have charge $+2/3$ and $-1/3$ respectively.
A non-zero charge density $Q/V$ is imposed by introducing 
chemical potentials for the bosons and fermions.
In chemical equilibrium, the fermions and bosons chemical potential are related, $\mu_B = 2 \mu_F$, so we concentrate on the bosons. For fixed external 
charge the chemical potential and the temperature 
are not independent, $\mu_B = \mu_B(T)$. Usually, if the bosons 
were massive in the non-interacting limit, they would undergo Bose-Einstein 
condensation at some critical temperature $T_{BC} \propto (Q/V)^{2/3}/m_B$ 
where $m_B$ is the boson mass. Operationally, this is the temperature at which 
the boson chemical potential $\vert \mu_B\vert = m_B$. Obviously, if 
$m_B \rightarrow 0$, the charge is always in the zero momentum condensate 
i.e. $T_{BC} \rightarrow \infty$. Interactions change this  
as the bosons get a mass $m_B(T) \sim g T$ in a thermal bath,
but the reasoning is essentially the same with  $\vert \mu_B\vert = m_B(T)$ 
\weldon. Now 
consider our non-commutative theory. Below $T_0$ the theory has a behaviour 
very much analogous to a conventional bosonic field theory at finite 
temperature. Above $T_0$ on the other hand, the minimum of the dispersion 
relation is at non-zero transverse momentum. Bosons with such a 
dispersion relation cannot undergo Bose-Einstein condensation. This is 
because the chemical potential is bounded by the minimum of the boson 
dispersion relation, $\vert \mu_B(T) \vert \leq \omega (p_\perp^{min},T)$.
For $\vert \mu_B(T) \vert \rightarrow \omega (p_\perp^{min},T)$  the integral 
for the global charge density $Q/V$
\eqn\pole{
Q/V \sim \int d^3p {1\over \exp\beta(\omega(p)- \omega(p^{min}_\perp)) -1} \sim 
\int dp_\perp \, {p_\perp\over |p_\perp - p^{min}_\perp|}
}
diverges when $p_\perp^{min} \not= 0$.\foot{
That the dispersion relation prevents the condensation of zero modes is analogous to the 
effect discussed in 
\gubser . However here the chemical potential is not an independent parameter that can be 
varied to go from the disordered to the ordered phase, but a constraint which adjusts itself 
so as to insure charge conservation. } 
We thus expect that, at high densities $T_0$ is the maximal critical temperature for Bose-Einstein
condensation.

\newsec{Gauge Theories at Finite Density}

We have studied the behaviour of non-commutative 
theories in a thermal bath. Temperature allowed us to probe the theory at 
different scales and observe the deviations with respect to the ordinary 
behaviour. There are however other ways of probing the system. 
It is interesting to analyse how much the effects we have obtained depend 
on the fact of having a thermal bath or are reproduced in other physical 
settings. In particular, we can consider theories at finite density.
The effects of non-vanishing chemical potential have been studied in 
\gll\ in a non-relativistic, non-commutative field theory at finite 
temperature. It turned out that the effects of non-commutativity depend 
strongly on the value of the chemical potential. In the previous section 
we have analysed the Wess-Zumino model at finite temperature and density. 
There again the behaviour of the system with respect to Bose-Einstein
condensation depended strongly on the density. We want to study now
a situation where the only scale present is provided by the chemical
potential, namely finite density at zero temperature.  

Let us take as an example ${\cal N}\!=\!2$ non-commutative gauge theory.
The spectrum consists of a vector field, two Weyl spinors and a complex scalar.
Ordinary ${\cal N}\!=\!2$ Yang-Mills theory has a non-anomalous $SU(2)_R$ 
symmetry under which vector and scalars are singlets and the two 
${\cal N}\!=\!2$ gluinos transform as a doublet. This is still valid for 
the non-commutative version. There is a $U(1) \in 
SU(2)_R$ under which the gluinos $\lambda$ and $\psi$ transform with 
charges $1$ and $-1$ respectively. 
A non-zero density of the associated conserved charge is imposed by 
introducing a chemical potential $\mu$. Positive chemical 
potential corresponds to an excess of $(\lambda,{\bar \psi})$ quanta over 
$(\psi,{\bar \lambda})$ quanta in the system, and vice-versa for negative 
chemical potential. To be definite we consider $\mu$ positive.

\ifig\waves{$\omega$ as a function of $p_\perp$
for different values of the chemical potential ($p_3=0,g=0.1,\theta=1$).}
{
\epsfxsize=3.8truein\epsfysize=2.4truein
\epsfbox{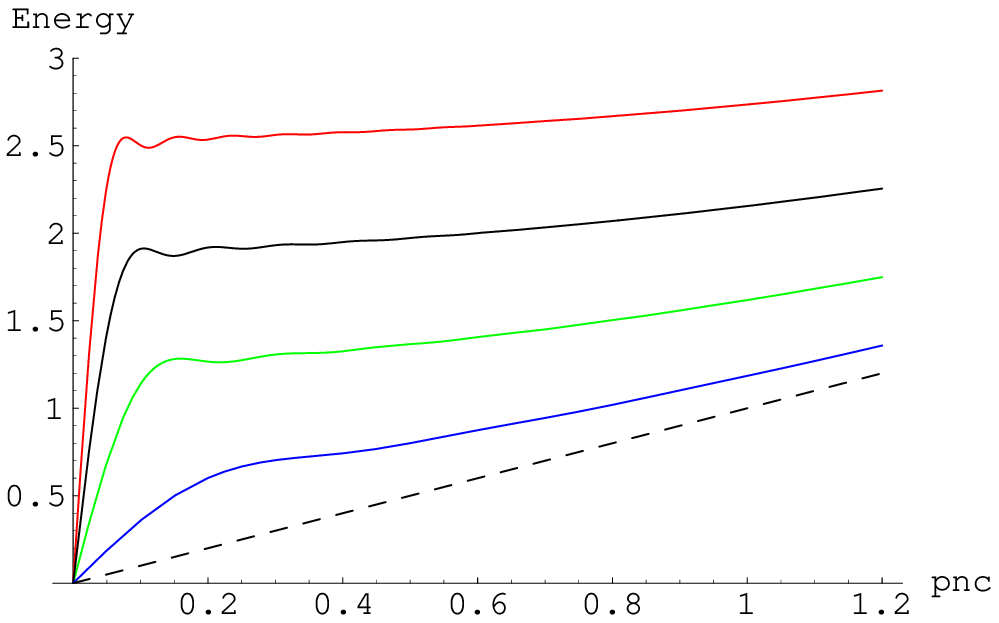}
} 

For simplicity we will consider a $U(1)$ ${\cal N}\!=\!2$ non-commutative
theory. The one loop contribution the the scalar self-energy due to 
a finite fermion density is given by
\eqn\dens{
\Sigma = 8 g^2 \int {d^3 k \over (2 \pi)^3} {{\rm sin}^2 
{{\tilde p} k \over 2} \over k} \Theta(\mu -k) \, ,
}
where $\Theta(x)$ is the step-function.
We obtain the following modified dispersion relation
\eqn\energyD{
\omega^2= p^2  \, + \, {g^2 \mu^2 \over \pi^2} \, - \, 
{4 g^2 \over \pi^2  {\tilde p}^2} 
\, {\rm sin}^2  {\mu {\tilde p} \over 2}.
}

The dispersion relation \energyD\ is similar to the one obtained
for ${\cal N}=4$ gauge theories in a thermal bath, with the chemical potential
$\mu$ playing the role of temperature. The spectrum is gapless as can be 
seen in \waves. For small transverse momentum we 
have group velocities bigger than one,
\eqn\biggercD{
\omega^2 = \left( 1+ {g^2 \theta^2 \mu^4 \over 12 \pi^2} \right) {p_\perp}^2.
}
For chemical potential $\mu^2>>{1 \over g \theta}$ and $p_\perp > p^0_\perp$,
\eqn\plimit{
p_\perp^0 \approx \left( {2 g^2 \mu \over \pi^3 \theta} \right)^{1 \over 3}
\, ,
}
the scalar modes have a mass $m\approx {g \mu \over \pi}$.
A new feature of \energyD\ with respect \dispersionnfour\ is
the appearance of a very dispersive region at 
${2 \pi \over \mu \theta} <p_\perp<p_\perp^0$. In this region the 
$sin^2$ term on \energyD\ dominates and 
$\omega$ has a series of maxima and minima
with period ${2 \pi \over \mu \theta}$. The amplitude of the 
oscillations grows with the density, but they are always of
small relative size with respect to the first maximum.

\newsec{Discussion}

We have investigated the dispersion relations for scalar excitations in
non-commutative supersymmetric
theories at finite temperature (and density). We found several striking
new phenomena. In particular, it
seems that wavefronts can travel faster than the speed of light.
Also, in the
Wess-Zumino model, above some 
temperature, there is a dip in the dispersion relation and the minimum of
energy is at non zero 
transverse momentum. 

These features are in essence already present in
noncommutative theories in 
vacuum and we expect them to be quite generic consequences of spatial
non-commutativity. At one loop, non-planar diagrams give rise to
pole-like
divergences in the dispersion relations and to the
now famous issue of UV/IR mixing. 
While in supersymmetric theories these puzzling divergences 
are absent in vacuum, they reappear in disguise  in presence of a
thermal bath.  Finite temperature or density not only breaks 
supersymmetry but also 
provides a natural cut-off which allows to investigate the issue of UV/IR
mixing in a controlled setting.
In particular, the new effects we observed are more pronounced as we
increase 
temperature or a chemical potential. 
Since these parameters act as UV-regulators our findings are a direct
consequence of the UV/IR mixing in non-commutative field theories.

We have found dispersive
wavefronts
travelling faster than the speed of light.
As we tried to argue this is perhaps not
inconsistent since these theories are not
Lorentz invariant.
The \nfour\ theory can be realized as the decoupling
limit of a D3-brane in a B-field background in string theory.
Since the string theory is Lorentz invariant it is a bit puzzling
how velocities faster than the speed of light can appear.
One possibility is that the tower of additional string states
modifies the dispersion relation in a convenient way. This question  
is currently under investigation.

We have limited ourself to one loop, real tadpole
corrections. 
At two loop, we expect  scalar excitations in a thermal bath to be also
unstable,
both because of Landau damping and particle pair creation.
While these effects are subleading in $g$, 
the instabilities will also affect the long time behaviour of 
the propagating wavefront. It would be interesting to investigate this
point further.

\vskip1cm

\centerline{\bf Acknowledgments}
\vskip2mm
We would like to thank L. Alvarez-Gaum{\'e}, J. Barb{\'o}n, J. Gomis, 
C. Manuel for discusions.

\listrefs
\end